\documentclass[aip,apl,twocolumn,amsmath,amssymb]{revtex4}
\usepackage{bm}
\usepackage{amssymb}
\usepackage{amsmath}
\usepackage{latexsym}
\usepackage[dvips]{graphicx}
\begin{document}
\author{Stanislav Stoupin}
\email{sstoupin@aps.anl.gov}
\affiliation{Advanced Photon Source, Argonne National Laboratory, Lemont, Illinois 60439, USA}

\title{Self-detection of x-ray Fresnel transmissivity using photoelectron-induced gas ionization}

\begin{abstract}
Electric response of an x-ray mirror enclosed in a gas flow ionization chamber was studied under the conditions of total external reflection for hard x-rays. It is shown that the electric response of the system as a function of the incidence angle is defined by x-ray Fresnel transmissivity and photon-electron attenuation properties of the mirror material. A simple interpretation of quantum yield of the system is presented.  The approach could serve as a basis for non-invasive in-situ diagnostics of hard x-ray optics, easy access to complementary x-ray transmissivity data in x-ray reflectivity experiments and might also pave the way to advanced schemes for angle and energy resolving x-ray detectors. 
\end{abstract}


\maketitle



Electric self-detection of x-ray-induced photoemission from an object can be considered as non-invasive monitoring of the radiation flux because optimization of detection of the generated electric carriers is focused on creating efficient charge collection in the exterior of the object.
In the soft x-ray regime (photon energies $<$~5 keV) photoemission-based self-detection of x-ray flux is readily performed in high vacuum environment. An object (usually conductive) is in direct contact with a conductive holder that is connected to the electrical ground through a current meter. 
As an uncompensated charge develops due to escape of photoelectrons a compensating electric current flows to the sample holder and is registered by the current meter.  The magnitude of this current can serve as a measure of the incident or absorbed photon flux, thus offering non-invasive monitoring capability and a probe of x-ray absorption. This measurement mode is often referred to as total electron yield since all electrons that exit the surface are detected, independent of their energy (e.g., \cite{deGroot_book,Ebel04,Vlachos04}). 

\begin{figure}[h]
\centering\includegraphics[width=0.5\textwidth]{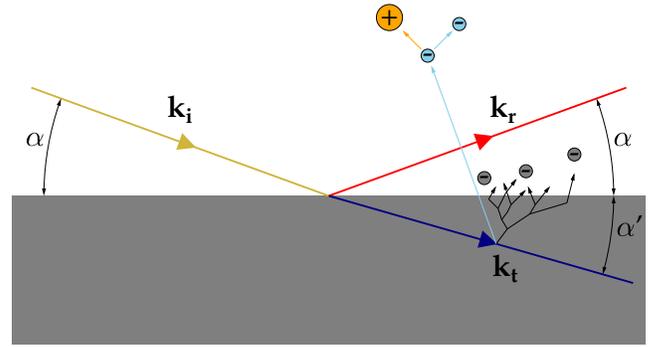}
\caption{Grazing incidence geometry illustrating the incident, the transmitted and the reflected wave. The transmitted wave generates ionization events in the mirror material due to x-ray absorption. The escaping fast photoelectons (light blue) ionize the surrounding gas while the low-energy secondary photoelectrons (gray) do not participate in this ionization process.}
\label{fig:gzi}
\end{figure}

The same self-detection approach has been utilized for collection of x-ray absorption spectra in the hard x-ray regime (e.g., \cite{Martens78,Martens79,Erbil88}). The core-level x-ray absorption edge manifests itself by an increase in the electric current due to cascades of secondary electrons caused by generation of Auger electrons and the resulting x-ray absorption spectra probe sample depths $L\approx$~1000~$\rm \AA$. 
For hard x-rays the detection technique can take advantage of negligible x-ray absorption in a light gas environment such as helium \cite{Kordesch84, Guo85}. Instead, helium is subject to efficient ionization by fast photoelectrons escaping the material (Fig.~\ref{fig:gzi}), which provides enhancement in the quantum detection yield. The same approach is used in conversion electron M{\"{o}}ssbauer spectroscopy (e.g., \cite{Jones78}). 

Hard x-ray flux monitoring on optical elements in a single electrode configuration has been demonstrated and studied earlier \cite{pat:Stoupin13,Stoupin_arxiv14,Stoupin_SPIE14}. Weak signal contrast for x-ray beam being on/off the optical element was attributed to ionization of surrounding air. In this work, self-detection of hard x-ray induced photoelectrons is used in grazing incidence geometry to probe Fresnel transmissivity of an x-ray mirror enclosed in a flowing helium gas. The measured quantum detection efficiency as a function of the grazing angle is described by a simple model which takes into account the photon energy of the incident beam, properties of the mirror material and ionization properties of the gas. The results obtained using the simple electric measurement procedure could be used for accurate assessment of the critical angle in total external reflection without detection of the reflected beam, to indicate changes to the state of the surface and to estimate the effective photoelectron escape depth. Thus, simple electric self-detection measurements could serve as a basis for non-invasive in-situ diagnostics of hard x-ray optics and provide access to complementary x-ray transmissivity data in x-ray reflectivity experiments. The self-detection approach using photoemission might also pave the way to advanced schemes for angle/energy resolving x-ray detectors.  

The problem to be addressed falls into the domain of grazing incidence x-ray photoemission spectroscopy (GIXPS), which was established by Henke~\cite{Henke72} as a method capable of determination of material constants and surface characterization. Further developments were performed \cite{Fadley74} including generalization to multilayer structures \cite{Chester93} followed by experimental effort \cite{Kawai95,Hayashi96}. The studies were mostly concentrated in the soft x-ray domain dictated by the small radiation penetration depth for applications in surface science. The variation of photoelectron yield in grazing incidence with photon energy in the hard x-ray domain was studied ~\cite{Martens78}. It was established that the fine structure in the photoelectron yield as a function of the photon energy agrees with the fine structure of absorption spectra (EXAFS). 
In spite of such extensive developments, applications of hard x-ray GIXPS remain limited to date~\cite{Fadley05,Fadley10,Kawai10}. 

Specular reflection of electromagnetic waves is described by the Snell's law and the Fresnel equations (e.g.,\cite{Hau-Riege_book}).
As pointed out by Henke~\cite{Henke72}, the approximate x-ray reflection-refraction theory is adequate to describe photoelectron yield in the hard x-ray regime.
In the grazing incidence geometry (Fig.~\ref{fig:gzi}) with the grazing angle of incidence $\alpha \ll 1$ of the incident wavevector $\mathbf{k_i}$ 
and the grazing angle $\alpha' \ll 1$ of the transmitted wavevector $\mathbf{k_t}$ 
the amplitude of the reflected wave can be approximated as

\begin{equation}
r = \frac{\alpha - \alpha'}{\alpha + \alpha'} .
\label{eq:t}
\end{equation} 
The grazing angle $\alpha'$ can be found using the Snell's law

\begin{equation}
\cos{\alpha} = (1-\delta + i \beta)\cos{\alpha'} ,
\label{eq:snell}
\end{equation} 
where $\delta \ll 1$ and $\beta \ll 1$ are the real and the imaginary corrections to the refractive index of the mirror material $n = 1 - \delta + i\beta$.  
The refractive index of a light gas such as helium can be replaced with unity since the corrections to the refractive index for He in a flow chamber at close-to-normal conditions are several orders of magnitude smaller than those of the high-density mirror material. The penetration depth of the transmitted wave is given by  

\begin{equation}
\Lambda(\alpha) = \frac{1}{2 k Im(\alpha')},
\label{eq:Lam}
\end{equation} 
where $k = |\mathbf{k_i}| = |\mathbf{k_r}|$ is the absolute value of the wavevector and $Im(\alpha')$ is the imaginary part of $\alpha'$.
The number of photoelectrons created in the material at a depth $z$ per second, within an increment $dz$, is 

\begin{equation}
n^e(z,\alpha) =  \frac{F_0 (1-|r(\alpha)|^2)}{\Lambda(\alpha)} \exp{\left(-\frac{z}{\Lambda(\alpha)}\right)}dz .
\label{eq:nez}
\end{equation}
Here, $F_0$ is the number of incident x-ray photons per second (i.e, photon flux) and $(1-|r(\alpha)|^2)$ represents Fresnel transmissivity. 
The probability of photoionization is given by the penetration depth $\Lambda(\alpha)$ which is inversely proportional to the linear photoabsorption coefficient $\mu$. 
The dependence on the photon energy of the incident wave $E_X$ is omitted here for brevity.

A fraction of the excited electrons completely escapes the mirror material. On the way to the surface these electrons exhibit inelastic scattering events, which results in the reduction of their initial energies. The electrons can escape the material only if they are generated within a certain characteristic depth known as electron inelastic mean free path (IMFP), which is a function of the electron energy \cite{NIST71}. 
Inelastic scattering does not only reduce the energy of the primary electrons but also produces a cascade of secondary electrons with smaller energies. 
A simplified analytical description of such complicated process can be offered based on several assumptions arising from experimental observations as derived in \cite{Stohr_book}. An assumption is made that the energy distribution of low-energy secondary photoelectrons is independent on the primary electron energy once it is higher than about 20~eV and that the number of the secondary photoelectrons is proportional to the incident photon energy $E_X$. The number of photoelectrons generated per one x-ray absorption event is referred to as the electron gain factor $G^e$. 
In analogy to attenuation of x-rays a quantity $1/L$ is introduced as a linear electron-attenuation coefficient (where $L$ is the effective electron-energy-independent escape depth of photoelectrons) that describes the electron scattering process as an attenuation of a single primary electron multiplied by the gain factor $G^{e}$.

If the x-ray mirror is enclosed in a He flow chamber gas impact ionization events produced by the secondary photoelectrons can be neglected since the energy required to produce one ion pair is $W_{g} \simeq$~40.3~eV \cite{Weiss56} while the energies of the secondary electrons do not exceed $\approx$~20~eV \cite{Henke77}. In addition, absorption cross section for hard x-rays in He is negligible compared to the ionization cross section by photoelectron impact {e.g., \cite{Shah88}}. Thus, the electric carriers generated in the gas flow chamber originate from the photoelectric response of the mirror material. 

Electrons generated in the depth increment $dz$ contribute a fraction $dY(z,\Omega,\alpha)$ to the integral electron yield emitted into solid angle $\Omega$

\begin{equation}
dY(z,\Omega,\alpha) = \frac{\Omega}{4 \pi} G^e n^e(z,\alpha) \exp{(-z/L)}.
\label{eq:dy}
\end{equation}

Here an isotropic angular distribution of the photoelectrons is assumed. 
In this approximation it is expected that detection of the integral electron yield (without discrimination in the electron energy and the electron emission direction) and the random  gas impact ionization process provide averaging of the spatial distribution of charge carriers and the net effect is consistent with isotropic distribution of photoelectrons. An improvement to this approach can be considered based on more detailed theories of x-ray-induced photoemission (e.g.,~\cite{Chester93}). 
 
The electron yield $Y(\alpha)$ can be found by integrating Eq.~\ref{eq:dy} over the thickness $z$ and the solid angle $\Omega$.

\begin{equation}
Y (\alpha) = \frac{1}{2} F_0  G^e (1-|r(\alpha)|^2) \frac{L}{\Lambda(\alpha) + L}.
\label{eq:y}
\end{equation} 

It is convenient to attribute the number of charge carriers $n_q$ generated by a single photoelectron in the gas flow chamber to the ratio of the maximum photoelectron energy $E_{pe} \simeq E_X$ and the ion pair production energy $W_{g}$. 

\begin{equation}
n_q = 2 \frac{E_X}{W_g}
\label{eq:n_q}
\end{equation} 

The quantum yield in the measurement of electric current is defined as the ratio of the number of detected charge carriers per second to the incident photon flux $F_0$. 
It is given by

\begin{equation}
Q (\alpha) = \frac{1}{2} \epsilon_q n_q G^e (1-|r(\alpha)|^2) \frac{L}{\Lambda(\alpha) + L},
\label{eq:Q}
\end{equation} 
where $\epsilon_q$ is the charge collection efficiency. In Eqs. \ref{eq:dy},\ref{eq:y},\ref{eq:Q}  by analogy to \cite{Stohr_book} $L$ is the effective photoelectron escape depth and $G^{e}$ is the electron gain factor. Since the secondary electrons are not expected to contribute to the electric response these parameters can be attributed to fast gas-ionizing photoelectrons with energies $E_{pe} \gg W_{g}$.

\begin{figure}[h]
\centering\includegraphics[width=0.45\textwidth]{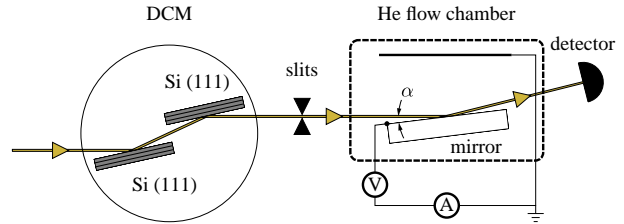}
\caption{Experimental setup for self-detection of Fresnel trasmissivity using photoelectron-induced gas ionization. Synchrotron bending magnet radiation is monochromatized using double-crystal monochromator (DCM). The monochromatized beam is shaped using slits and is incident on a Pd mirror placed in a He flow chamber. The metallic surface of the mirror serves as the first electrode for collection of generated charge in the flow chamber. The second electrode is placed above the mirror. The electric current in the circuit is measured using a source meter which supplies a bias voltage. A solid state detector is placed downstream the mirror to measure the photon flux.}
\label{fig:setup}
\end{figure}

To test the theory described above a palladium x-ray mirror was placed on a non-conductive surface and contained in a helium flow chamber having an entrance and exit windows made of Kapton$^{\textregistered}$ film. The thickness of the Pd film was 230 $\mathrm{\AA}$. It was deposited on top of a 50-angstrom-thick layer of Cr on a polished Si substrate. The length of the mirror was 100~mm and the width was 20~mm. The metallic working surface of the mirror served as the first electrode while a separate second grounded electrode was placed inside the chamber above the mirror at a distance of about 20 mm. The experimental setup is shown in Fig.~\ref{fig:setup}. 

The experiment was performed at 1-BM beamline of the Advanced Photon Source. The x-ray beam incident on the mirror was delivered by a Si (111) double-crystal monochromator (DCM) which was slightly detuned to suppress high-order reflections at high photon energies. The electric current between the electrodes in the flow chamber was measured using a source meter with applied bias voltages of $\pm$ 200 V. It was verified that under these conditions the system operated in the regime of ionization chamber where nearly all generated electric carriers were being collected (i.e., $\epsilon_q$~=~1). The observed maximum electric currents were up to 10 nA while the current in the absence of x rays (i.e., dark current) was $\approx$~100~pA. The root mean square fluctuations in the measured signal were about 20~pA. A calibrated solid state detector was placed behind the flow chamber to measure the reflected x-ray flux (mirror in the beam) and the incident x-ray flux (mirror is out of the beam).  
The size of the incident beam was set to $0.02\times5.0$~mm$^2$ (vertical$\times$horizontal) using x-ray slits placed upstream of the mirror chamber. 
The incident beam was centered on the surface of the mirror. Simultaneous measurement of the reflectivity and the electric current were performed while scanning the mirror's grazing angle $\alpha$ at different photon energies selected by the double-crystal monochromator. Prior to each scan the incident photon flux was measured using the solid state detector. The measured values of the incident photon flux were in the range $1 \times 10^{8} - 1 \times 10^{9}$ photons/s. 

No difference in the absolute value of the electric current was found upon switching polarity of the bias voltage within the experimental uncertainties. It should be noted that the difference is expected in the presence of Auger photoelectrons with energies sufficient for impact ionization of the gas atoms and comparable with the applied bias voltage. The observation suggests that such Auger photoelectrons are not present in the photon energy range of this study (10 - 18~keV) for Pd. New types of Auger photoelectrons are expected when the photon energy crosses an absorption edge while Pd does not have absorption edges in the chosen range. 

\begin{figure}[h]
\centering\includegraphics[width=0.48\textwidth]{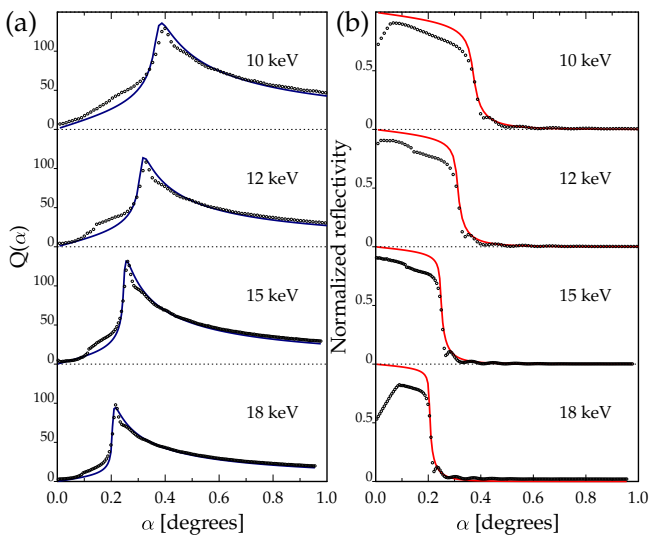}
\caption{(a) Electric response of Pd x-ray mirror in He flow ionization chamber as a function of the grazing angle of incidence at various incident photon energies. The experimentally measured electric response is shown by circles in the units of quantum yield (Eq.~\ref{eq:u_qy}). The fits to the theoretical expression for the quantum yield (Eq.~\ref{eq:Q}) are shown by the solid (blue) lines. (b) Fresnel reflectivity as a function of the grazing angle of incidence at the corresponding incident photon energies. The experimentally measured normalized reflectivity is shown by circles while the theoretical reflectivity $|r(\alpha)|^2$ for a thick perfect Pd mirror is shown by the solid (red) lines.}
\label{fig:fits}
\end{figure}

The resulting experimental curves for different photon energies are shown in Fig.~\ref{fig:fits}(a) (circles) only for the positive bias voltage. 
The measured electric current $I(\alpha)$ was converted to quantum yield as 
\begin{equation}
Q(\alpha) = \frac{I(\alpha)}{q_e F_m},
\label{eq:u_qy}
\end{equation}
where $q_e$ is the elementary charge and $F_m$ is the incident photon flux. 
Remarkably, the experimental data expressed in the units of quantum yield reveal generation of up to about one hundred of charge carriers per one incident photon. This observation suggests that a quantitative angular/energy resolving detector for hard x-rays can be constructed based on the grazing incidence geometry. 

In the first step of data analysis reflectivity data for each photon energy were aligned with the theoretical reflectivity $|r(\alpha)|^2$ in order to determine the angular offsets from the absolute angular scale. The precision of this alignment was about 0.0025~deg, which was half the angular step size in the experiment. 
In the second step nonlinear squares fitting of the electric response to the expression for the quantum yield (Eq.~\ref{eq:Q}) was performed using $G^e$ and $L$ as variable parameters. Photon-energy-dependent refractive index corrections for Pd were obtained from \cite{CXRO}. 
In this step reflectivity of the thick Pd mirror was used which is an approximation where variations in the yield curve due to the layered structure of the mirror are not taken into account. A variable correction to the angular scale $d \alpha$ was introduced to test angular selectivity of the quantum yield curve. The fits are shown in Fig.~\ref{fig:fits}(a) by the solid lines and the values of the fitted parameters are given in Table~\ref{tab:param}. The complementary experimental reflectivity data are shown in Fig.~\ref{fig:fits}(b) by circles and the corresponding theoretical reflectivity 
$|r(\alpha)|^2$ is plotted by the solid lines. The discrepancy between the theoretical and experimentally measured reflectivity can be attributed to a rough surface as evidenced by decrease in reflectivity below the critical angle and the layered structure of the mirror as evidenced by the appearance of Kiessig fringes. 

\begin{table}
\caption{Parameters extracted from fitting of the measured electric response with the expression for the quantum yield 
(Eq.~(\ref{eq:Q})):
$G^e$ - electron gain factor,
$L$ - effective photoelectron escape depth,
$\delta \alpha$ - correction to the angular scale.}
\begin{ruledtabular}
\begin{tabular}{l c c c}
$E_X$ [keV]    & $G^e$       &$L$ [$\rm \AA$]      &$\delta \alpha$ [degrees] \\
\hline
10       &1.57(0.07)   &155(11)       & 0.007($5\times10^{-5}$)  \\         
12       &1.24(0.07)   &168(12)       & 0.003($3\times10^{-5}$) \\  
15       &1.10(0.04)   &265(13)       & 0.002($1\times10^{-5}$)   \\  
18       &0.58(0.02)   &466(24)       & 0.002($1\times10^{-5}$)  \\         
\end{tabular}
\end{ruledtabular}
\label{tab:param}
\end{table}

The values for $\delta \alpha$ were found to be small, which indicates that the critical angle can be determined from the electric response with precision on the order of the angular step size. The effective photoelectron escape depth $L$ increases with the photon energy since electrons of higher energies are generated having greater IMFP. At the same time, the effective electron gain $G^e$ decreases which implies that a larger fraction of the excited electrons are converted to slow secondary photoelectrons which do not contribute to gas ionization. 
The quality of the fits was found to be good ($R^2$-factors of about 1\%). The remaining discrepancies below the critical angle can be attributed to incompletely removed contamination of the incident beam with high-energy radiation harmonics (e.g., photon energy of $3 \times E_X$ originating from the Si 333 reflection of the double-crystal monochromator). The discrepancies in the vicinity of the critical angle and above can be related to the structure of the mirror.
In addition, it is expected that surface contamination will impede photoelectron emission, thus clean surfaces should be used for more precise measurements of  $G^e$ and $L$. At the same time, a change in the electric signal with time could be considered as a diagnostic probe which is sensitive to surface contamination and surface damage.
Finally, the reasonable agreement between the model and the experimental data suggests that Fresnel transmissivity can be deduced from the electric response provided that the material constants are known. The shape of the quantum yield curve could be used to gain insights on the surface morphology using an appropriate model for $|r(\alpha)|^2$. Similarly, if the surface morphology is known the electric response could be used to measure materials constants.

\begin{figure}[h]
\centering\includegraphics[width=0.48\textwidth]{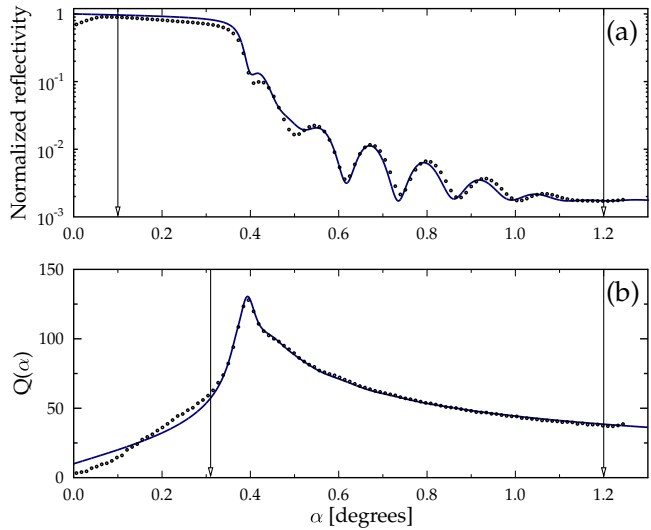}
\caption{(a) Experimentally measured reflectivity (open cirles) at 10 keV and fit (solid line) using Parratt's recursive method taking into account the layered mirror structure. (b) The obtained reflectivity function is used to fit the experimental quantum yield curve at 10~keV (open circles). The resulting fit (solid line) is of much better quality compared to the fit using the reflectivity of the thick mirror (Fig.~\ref{fig:fits}(a)). The vertical arrows indicate the angular ranges used in the fits ((a) and (b)).}
\label{fig:prt}
\end{figure}

In order to test these ideas, the layered structure of the mirror was taken into account by fitting the reflectivity using Parratt's recursive method \cite{Parratt54,Windt98}.
The resulting fit shown in Fig.~\ref{fig:prt}(a) confirmed the nominal thicknesses of the Pd film and the Cr layer. It also revealed Pd entrance surface roughness of about 9~${\mathrm \AA}$. The obtained reflectivity function was used to fit the yield curve using Eq.~\ref{eq:Q} at a photon energy of 10~keV. At this lower photon energy the photon-electron attenuation properties of the layered structure are dominated by the contribution of the Pd film. Thus, the penetration depth of the wavefield can still be approximated with that of a thick Pd mirror (Eq.~\ref{eq:Lam}). 
The resulting fit of much better quality is shown in Fig.~\ref{fig:prt}(b) (R$^2$-factor $<$~0.1~\%). The variable fit parameters were further refined ($G^e$ = 1.27(0.01), $L$ = 156(3) ${\mathrm \AA}$, $\delta \alpha$ = 8(1)~$\times$~10$^{-5}$ deg.). Also, a variable background was introduced to improve the fit quality (the background value obtained from the fit was $Q_0$~=~9.5(0.4)). 
It should be noted that the shoulder at angles above the yield peak position is now reproduced as well as minor oscillations that follow (i.e., Kiessig fringes in transmissivity). In fact, multilayer structures have been studied earlier using GIXPS. Observation of Kiessig fringes in the angular yield curves was reported \cite{Hayashi96} albeit in the soft x-ray region using high vacuum environment.     
It is also noted that absolute measurements of Fresnel transmissivity are less straightforward compared to conventional reflectance measurements (i.e., detection of the incident and the reflected x-ray flux using stand-alone x-ray detectors) since additional material parameters and related uncertainties are involved. A more detailed model of photoemission may be required to precisely decouple the structural response given by Fresnel transmissivity and the baseline signal due to angular dependence of the photon-electron attenuation factor.  

In summary, it is found that the electric response of the x-ray mirror enclosed in a gas flow ionization chamber can be described using a simple model for the quantum yield defined by x-ray Fresnel transmissivity and photon-electron attenuation properties of the mirror material. 
In the simple experimental arrangement accurate assessment of the critical angle in total external reflection can be performed without detection of the reflected beam. Also, it is proposed that from analysis of the obtained angular yield curve one can gain practical insight on the state of the surface and to estimate photoelectron escape depth. Thus, simple electric self-detection measurements of photoemission can provide non-invasive in-situ diagnostics of hard x-ray optics and access to complementary x-ray transmissivity data in x-ray reflectivity experiments. Furthermore, it becomes clear that electric measurements using photoelectron-induced gas ionization can be considered as a universal approach for non-invasive self-detection of x-ray flux in a wide variety of experimental arrangements. 

B. Shi is acknowledged for providing the x-ray mirror for the experiment. K. Goetze is acknowledged for the development of data acquisition software. M. Zhernenkov and Y. Cai are acknowledged for encouraging discussions on the possible applications.
Use of the Advanced Photon Source was supported by the U. S. Department of Energy, Office of Science, under Contract No. DE-AC02-06CH11357.


\end{document}